\documentclass[twoside,12pt]{article}
\usepackage{epsfig}

\def\Journal#1#2#3#4{{#1} {#2} (#4) #3 }

\def\NPA{{\em Nucl. Phys.} A}

\def\PLB{{\em Phys. Lett.} B}

\def\PRL{\em Phys. Rev. Lett.}

\def\PREP{\em Phys. Rep.}

\def\PRD{{\em Phys. Rev.} D}
\def\PRC{{\em Phys. Rev.} C}

\newcommand{\be}{\begin{equation}}
\newcommand{\ee}{\end{equation}}
\newcommand{\bea}{\begin{eqnarray}}
\newcommand{\eea}{\end{eqnarray}}

\begin{document}

\title{ \vspace{1cm}  Deep Core  muon neutrino rate and  anisotropy \\ by   mixing\\
and CPT violation }
\author{Fargion D.,$^{1,2}$ D'Armiento D.,$^2$\\
$^1$INFN, Rome University 1, Italy\\
$^2$Physics Department, Rome Univ. 1 Ple A.Moro 2, 00185, Rome, Italy\\
}
\maketitle
\begin{abstract}  Neutrinos are allowed to mix and to oscillate among their flavor. Muon and tau in particular oscillate at  largest values.
Last Minos experiment claimed \cite{14} possible difference among their matter and anti-matter masses, leading to a first violation of the most believed CPT symmetry.
 Isotropically born atmospheric muon neutrino at $E_{\nu_{\mu}}\simeq 20-80$ GeV, while up-going, they might be partially suppressed by mixing  in analogy to historical SuperKamiokande muon neutrino disappearance into tau, leading to large scale anisotropy signals.
   Here we show an independent   muon  rate foreseen  in Deep Core based on observed SK signals extrapolated to DeepCore mass and its surrounding. Our rate prediction  partially differ from previous  ones. The  $\nu_{\mu}$, $\bar{\nu_{\mu}}$ disappearance into  $\nu_{\tau}$,$\bar{\nu_{\tau}}$   is leading to a ${\mu}$, $\bar{\mu}$   anisotropy in vertical up-going muon track: in particular along channel $3-5$ we expect a huge rate (tens of thousand of events)  of neutral current events, charged current electron and inclined crossing muons. Moreover at channel  $6-9$ we expect a severe suppression of the rate due to muon disappearance (in CPT conserved frame). Such an anisotropy might be partially tested by two-three string detection at $E_{\bar{\nu_{\mu}}}\geq 45$. A CPT violation may induce a more remarkable suppression of vertical up-going tracks because of larger $\bar{\nu_{\mu}}$ reduction for $E_{\bar{\nu_{\mu}}}\geq 35$.

\end{abstract}
\section{Introduction}
The neutrino are very complex particles indeed. Their three light neutrino flavors mix in a complex way described by a
matrix born only in last few decades \cite{12},\cite{18},\cite{19}. Their presence maybe  recorded in small kiloton detector  or larger    (like SK and Super Kamiokande $22$ kiloton) ones.In largest size detectors as Icecube the higher characteristic $\nu_{\mu}$, $\bar{\nu_{\mu}}$ TeV  energy do not oscillate much and they do not exhibit  the  negligible oscillation along our narrow Earth. However the new born Deep Core, while tuning to few tens or even a few GeV  $\nu_{\mu}$, $\bar{\nu_{\mu}}$ energy, may hold memory of the $\nu_{\mu}$, $\bar{\nu_{\mu}}$ disappearance into tau. Many  have foreseen the $\nu_{\mu}$, $\bar{\nu_{\mu}}$ disappearance in Deep Core \cite{09}\cite{15}. We did use their prediction to calibrate the eventual CPT violation influence into their future rate \cite{00}. Here we review these predictions and now we reconsider our preliminary estimate, based on Super Kamiokande ones \cite{03a}, estimates that partially disagree with the previous results \cite{09}\cite{15}\cite{10}as well as \cite{09a}.  Deep Core,  is a new telescope or better a counter event  blurred at low energies (below $E_{\nu_{\mu}}\leq 30$ GeV)   because muons are tracing tracks mostly projected along one string: the inner cone within $\sim 30^{o}$ may contain any neutrino arrival direction around the string axis azimuth angle . At higher energy  $E_{\nu_{\mu}}\geq 45$ GeV the muon track, if inclined, may intersect two different  strings leading to a much accurate (a few degree) angular resolution. Therefore the DeepCore may test different energy regions at different degree of  resolution. Moreover most of the very inclined (respect the vertical) muons  (below $E_{\nu_{\mu}}\leq 30$ GeV) may intersect briefly the one string leading to a few (four-five) detector optical module (DOM) signal, in a very short timing clustering.  This mean that most of the inclined events may accumulate, as a noise, into the a few channel region that is at the same time the deposit of ten (or tens)  GeV shower, mostly Neutral Current, (NC), event by $\nu_{\mu}$, $\bar{\nu_{\mu}}$,$\nu_{\tau}$, $\bar{\nu_{\tau}}$ and  Charged Current, (CC), and NC due to electrons by $\nu_{e}$, $\bar{\nu_{e}}$. These few channel signals may also record  rare (nearly a thousand) tau appearance. However the previous noise will make difficult to discover tau appearance, while muon disappearance is still viable.
 \subsection{SuperKamiokande rate versus DeepCore}
\begin{figure}[htb]
\begin{center}
\epsfig{file=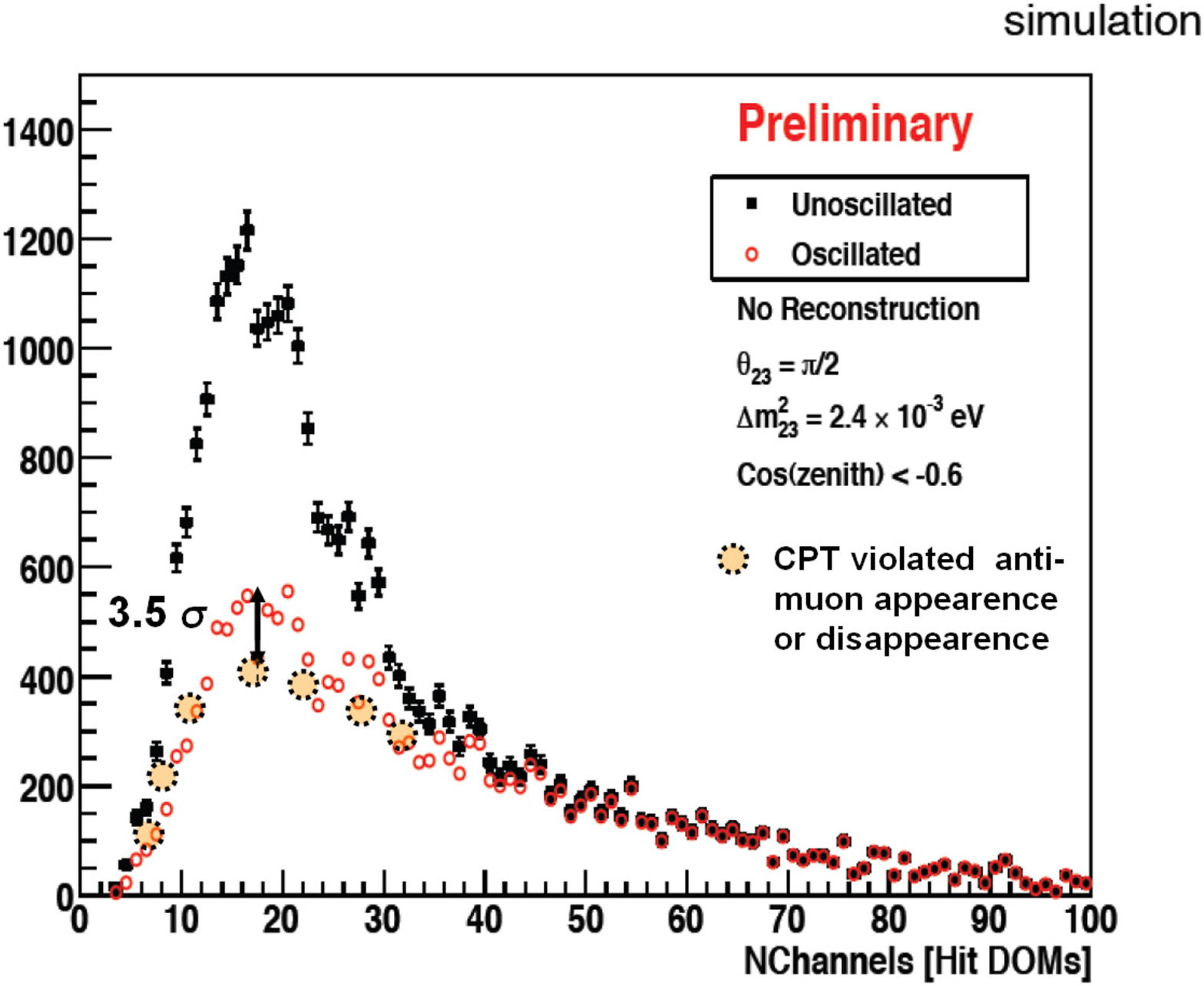,scale=0.13}
\epsfig{file=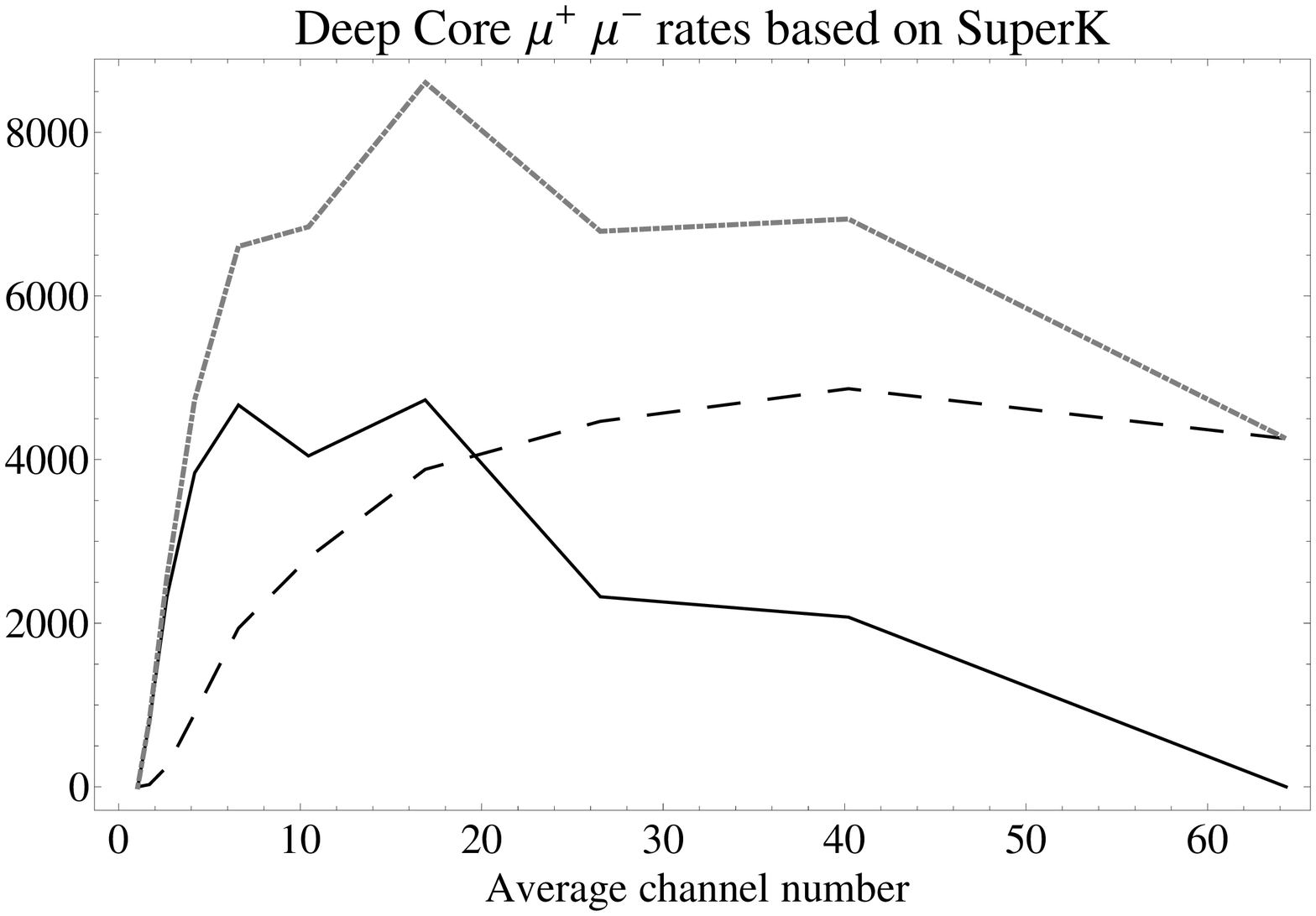,scale=0.25}
\epsfig{file=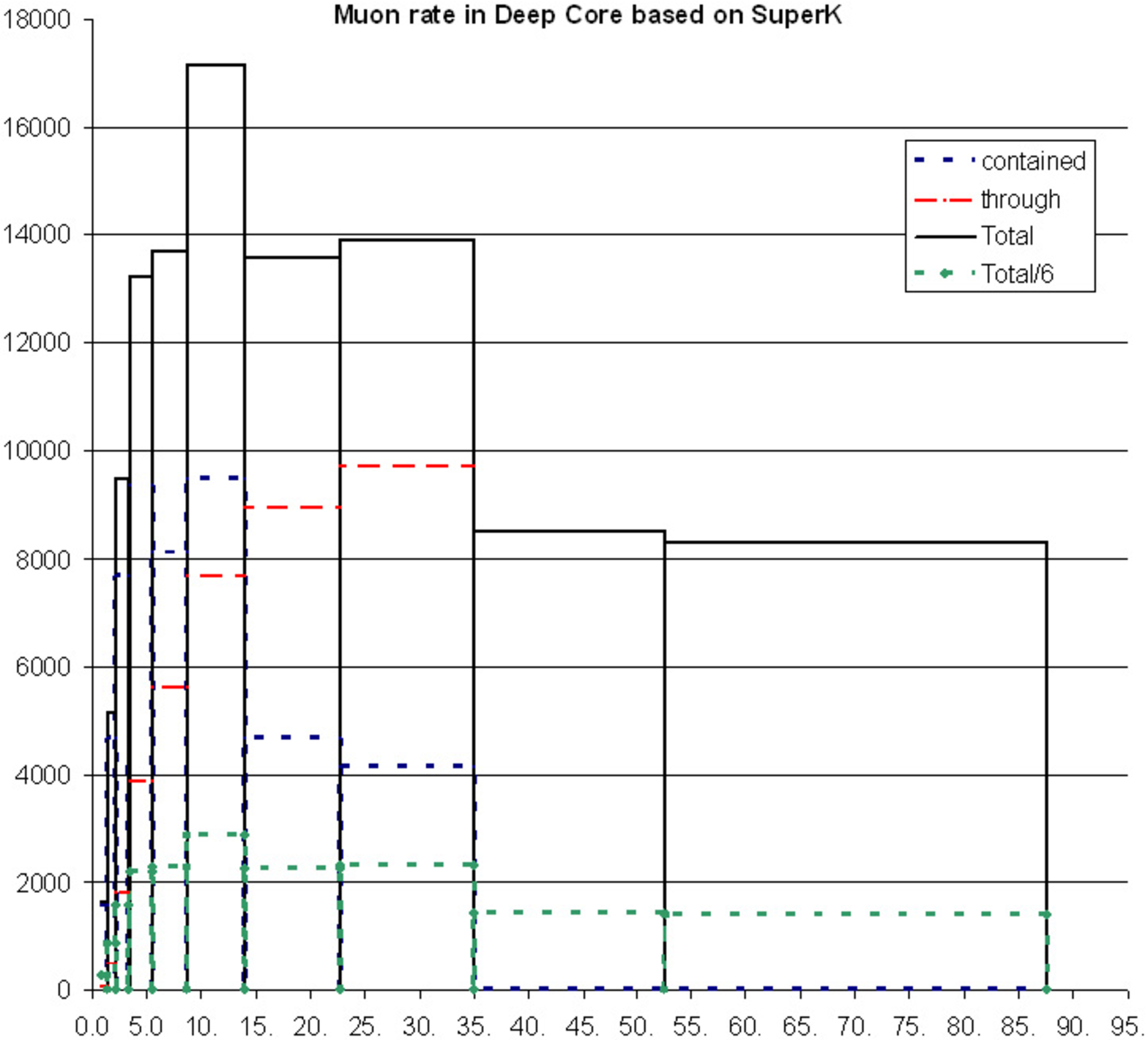,scale=0.22}
\caption{Left: The expected event rate following \cite{05},\cite{10}, of atmospheric muon neutrinos whose tracks are detected along the photo-tubes (DOM) channel.
The  expected ($\nu_{\mu}\rightarrow\nu_{\tau} $) \emph{and the anti-neutrino} muon ($\overline{\nu}_{\mu}\rightarrow\overline{\nu}_{\tau} $) induce a suppression. Our prediction for
 CPT violation, following \cite{00} is also shown. The SK event rate in Deep Core extrapolated from SK for each different nature (FC,PC,Upward stopping, Through going) are shown in the center :  dashed curve describe the upgoing stopping and upward through events; the FC and PC are described by lower continuous curve. The rate is much larger ($16$ times) probably because DeepCore is selecting only very vertical tracks, within $33^{o}$ from string axis, as shown in a normalized rate offered  in figure \ref{05}.}\label{123}
\end{center}
\end{figure}

 The simplest way to estimate the Deep Core  muon track and possible tau appearance (or better, the muon disappearance)   has been shown by Icecube MC simulation \cite{10}, as described in figure \ref{123}, with our additional  CPT violation expected influence\cite{00}. However we show here an independent derivation of the expected Deep Core rate, based on SuperKamiokande one \cite{01}. There are four main contribute to SK upgoing muons: Fully contained event (FC), born inside and decayed inside SK; partially contained event (PC), born inside but escaping outside the SK volume; the Upward Stopping Muons, born outside but decayed inside the detector; the upward through muon, just born outside , crossing and decaying again at external volume. Some care have been taken into account for the last Upward Stopping Muons and  upward through muon: the SK location is deeply surrounded by mountain rock, while Deep Core is within much less dense ice. Therefore we suppressed the two last rates by the density ratio ($\simeq 2.6$) to calibrate the expected rate in DeepCore , amplified by extrapolated volume ratio ($\frac{V_{DeepCore}}{22kT}$) versus Deep Core one at each energy range. Indeed the Deep Core volume is variable with the muon energy values due to photodetector thresholds and muon Cherenkov luminosity. We considered here the preliminary Deep Core effective volume variability following last Icecube articles \cite{05},\cite{10},\cite{15},\cite{20}. Our result is described in the right side of figure \ref{123},\ref{05} in linear scale along the channel number. We assumed an averaged neutrino  muon and anti-muon energy conversion, their length projected along the string at spread angle of $\theta\simeq  30^{o}$. The total event number derived by simplest SK-DeepCore translation is huge: $N \simeq 97.800$. Most of these events are not vertical but inclined. Therefore assuming a vertical beaming solid cone suppression (also to reconcile with Deep Core total expected rate) we selected  only those events within a cone angle  $\pm 33^{o}$, obtaining a fraction ($1- \cos\theta\simeq  0.16$) of the total rate, in this way now compatible with Deep Core preliminary global expectation $N \simeq 16.000$.
 \subsection{Rates and anisotropy}
\begin{figure}[htb]
\begin{center}
\epsfig{file=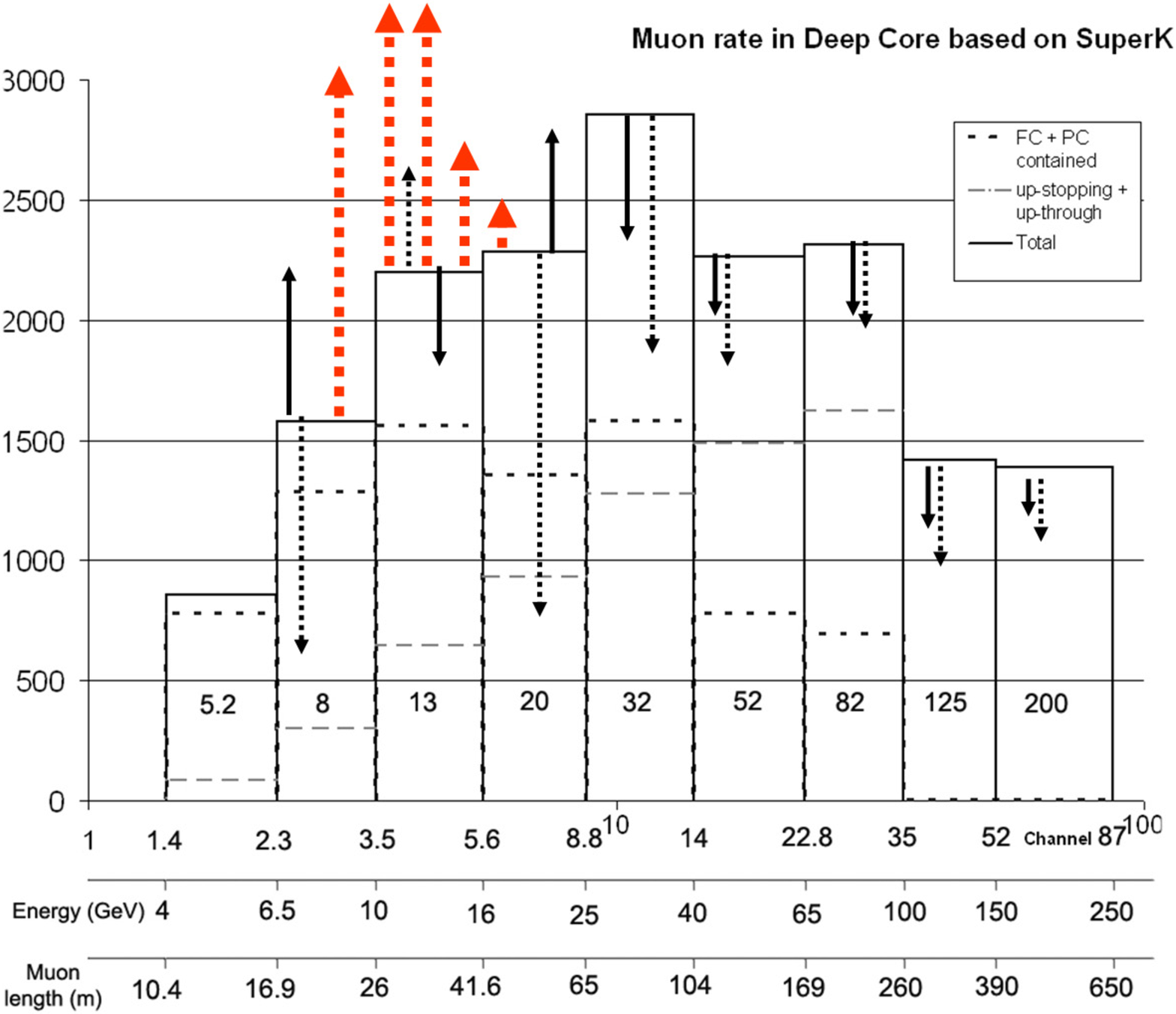,scale=0.24}
\caption{The expected event rate at each  muon (and anti-muon averaged) energy band, its corresponding muon (and anti-muon averaged) length, its corresponding channel number projected by a $\cos(\pi - \theta)$ factor; $\theta\simeq 33^{o}$.
 The rate are modulated by large red arrows standing for nearly one o two tens of thousand noise event, by a continuous black arrow due to CPT violated  influence, by a broken line due to CPT conserved muon tau disappearance. The influence of dashed arrow (muons) is twice larger than anti-muon ones because the twice larger neutrino versus antineutrino cross section. See Fig. \ref{613}. }\label{05}

 \vspace{1cm}

\epsfig{file=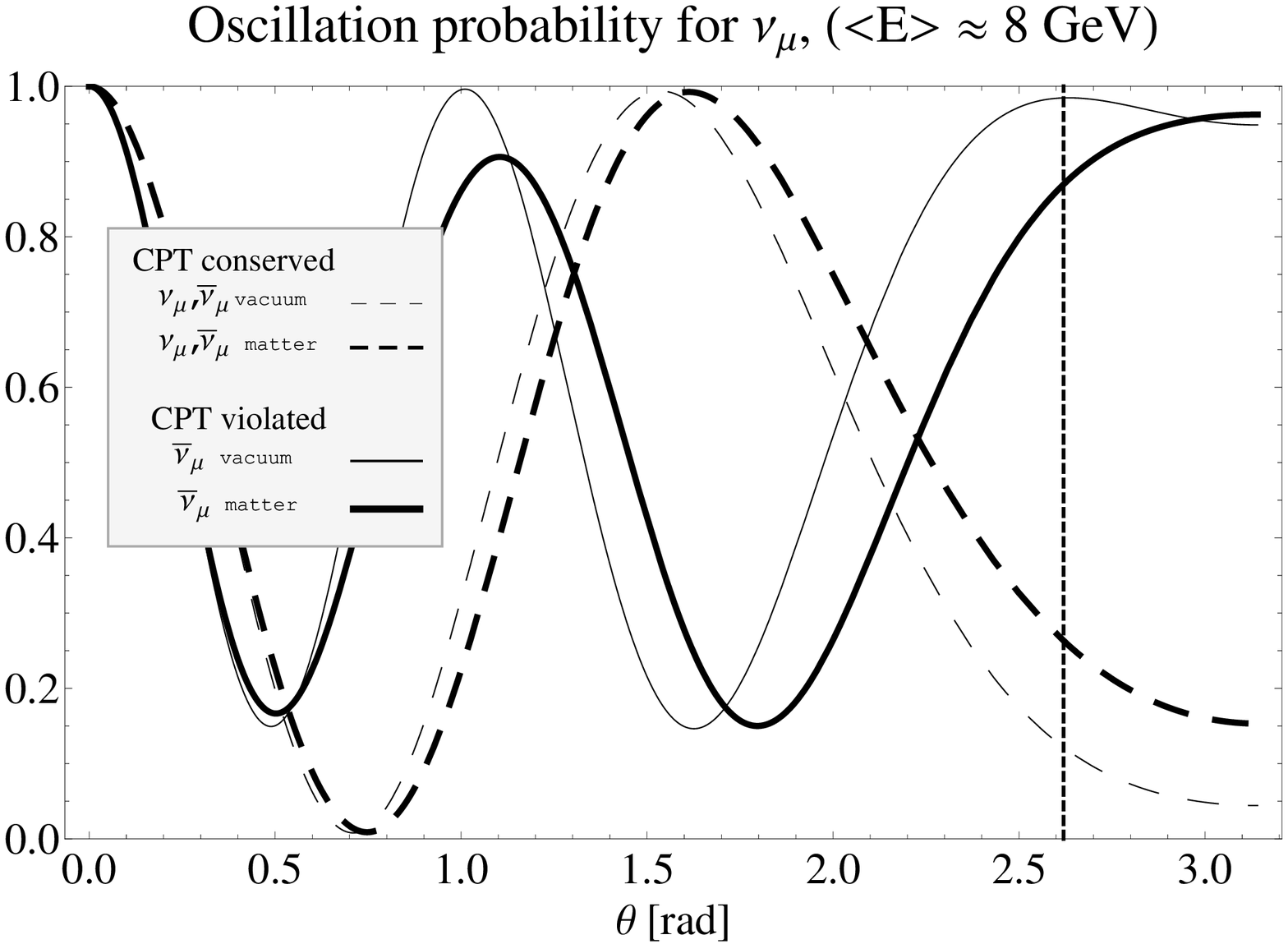,scale=0.22}
\epsfig{file=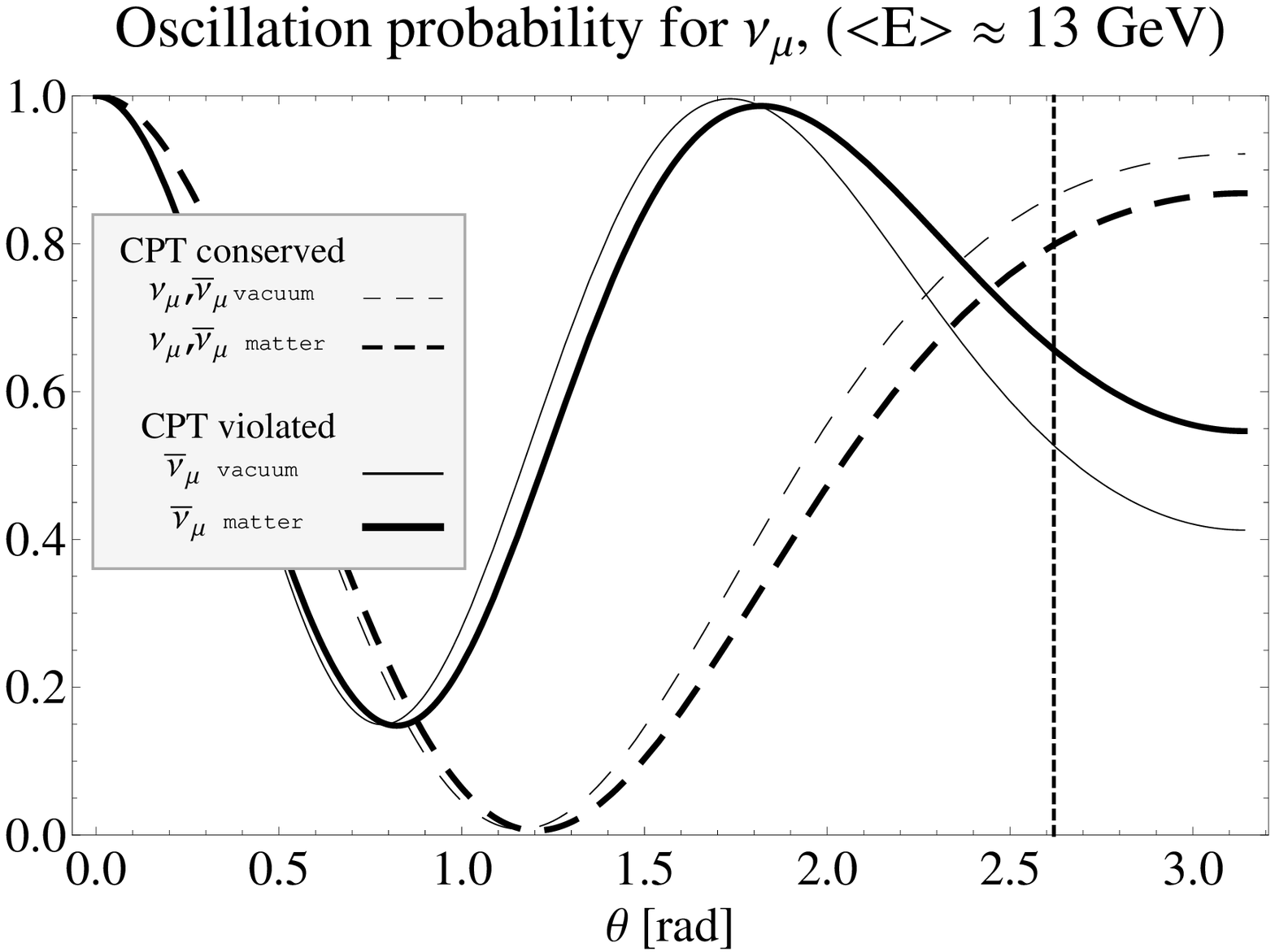,scale=0.22}
\epsfig{file=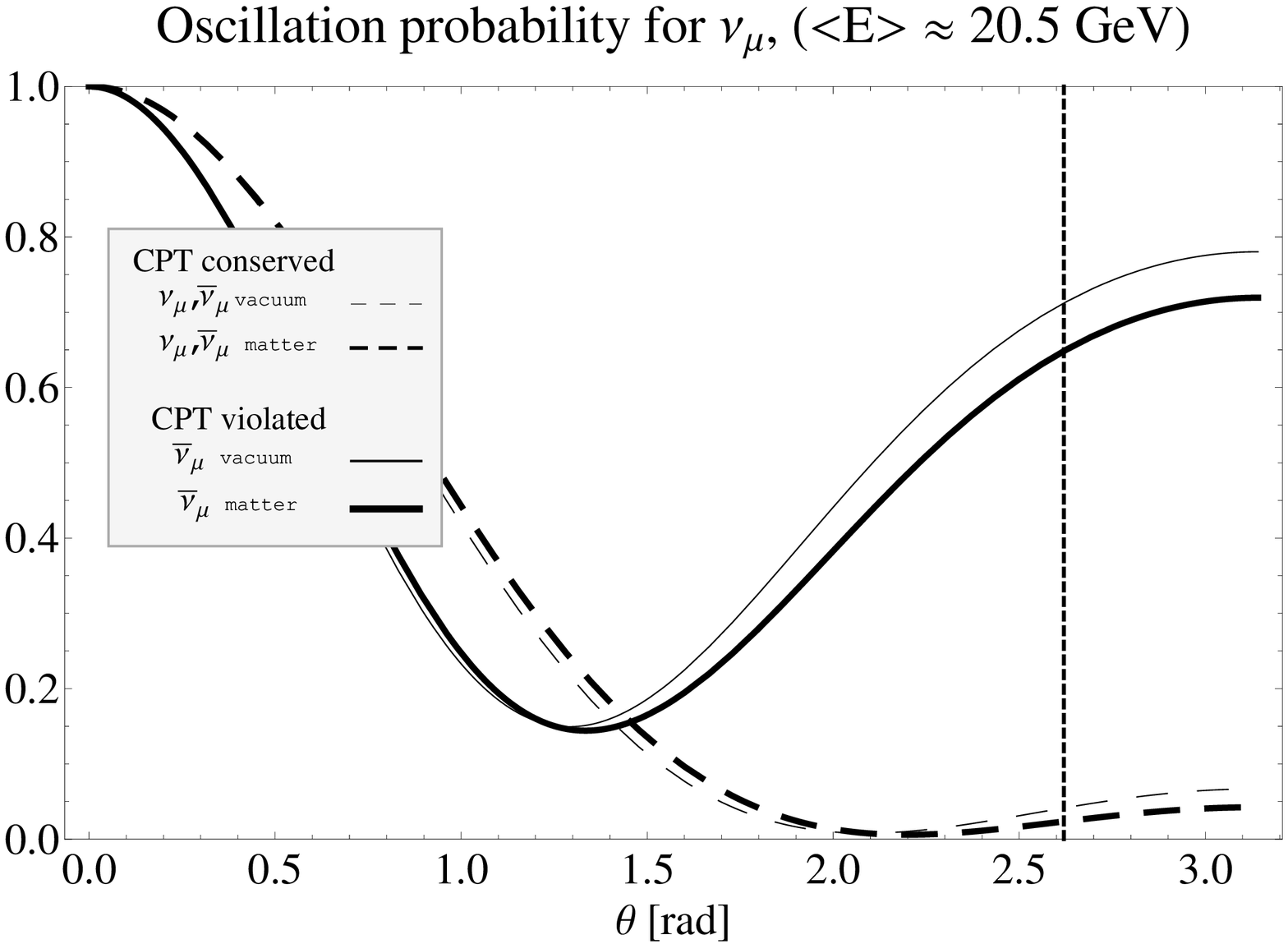,scale=0.22}
\epsfig{file=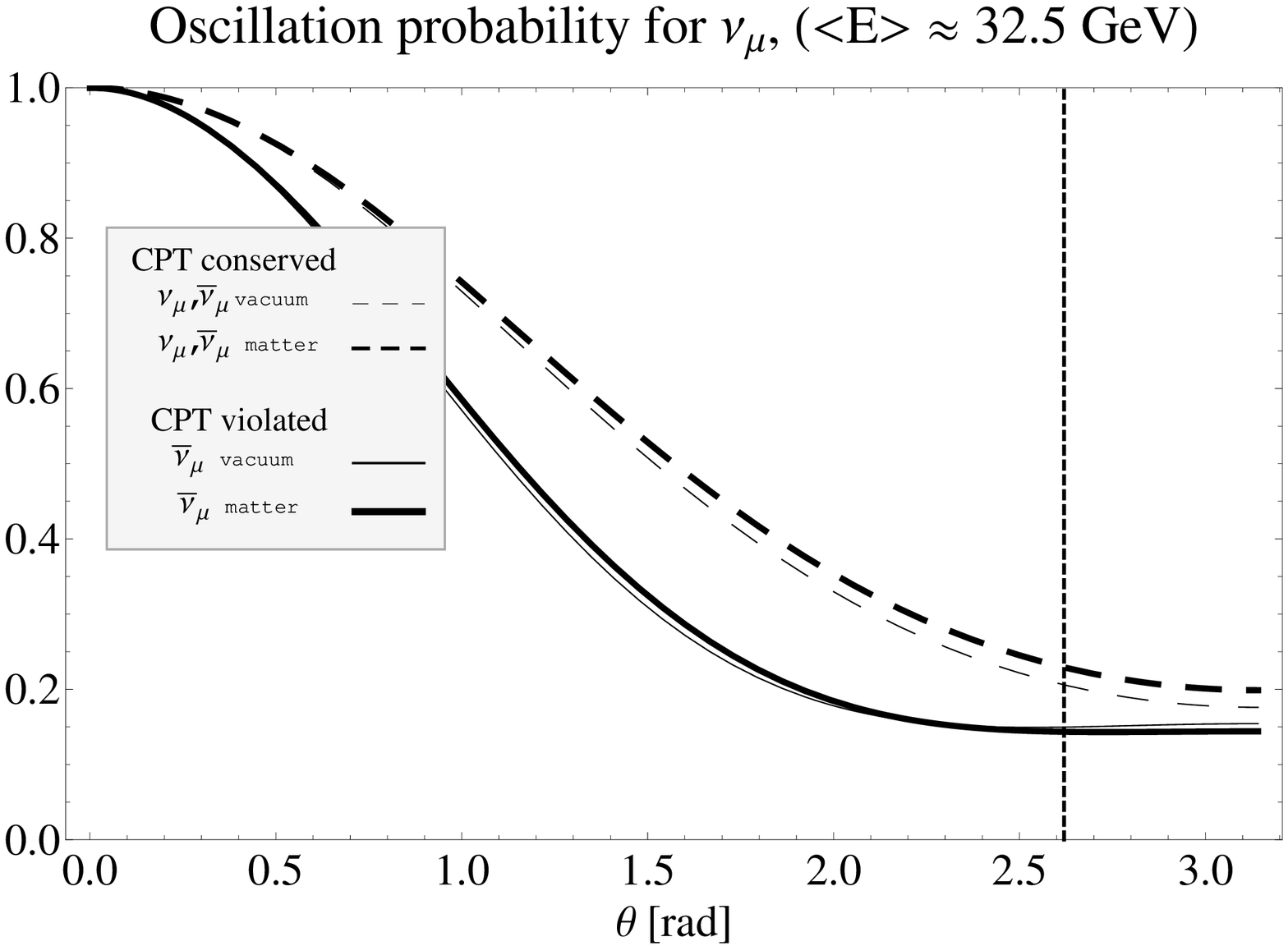,scale=0.22}
\epsfig{file=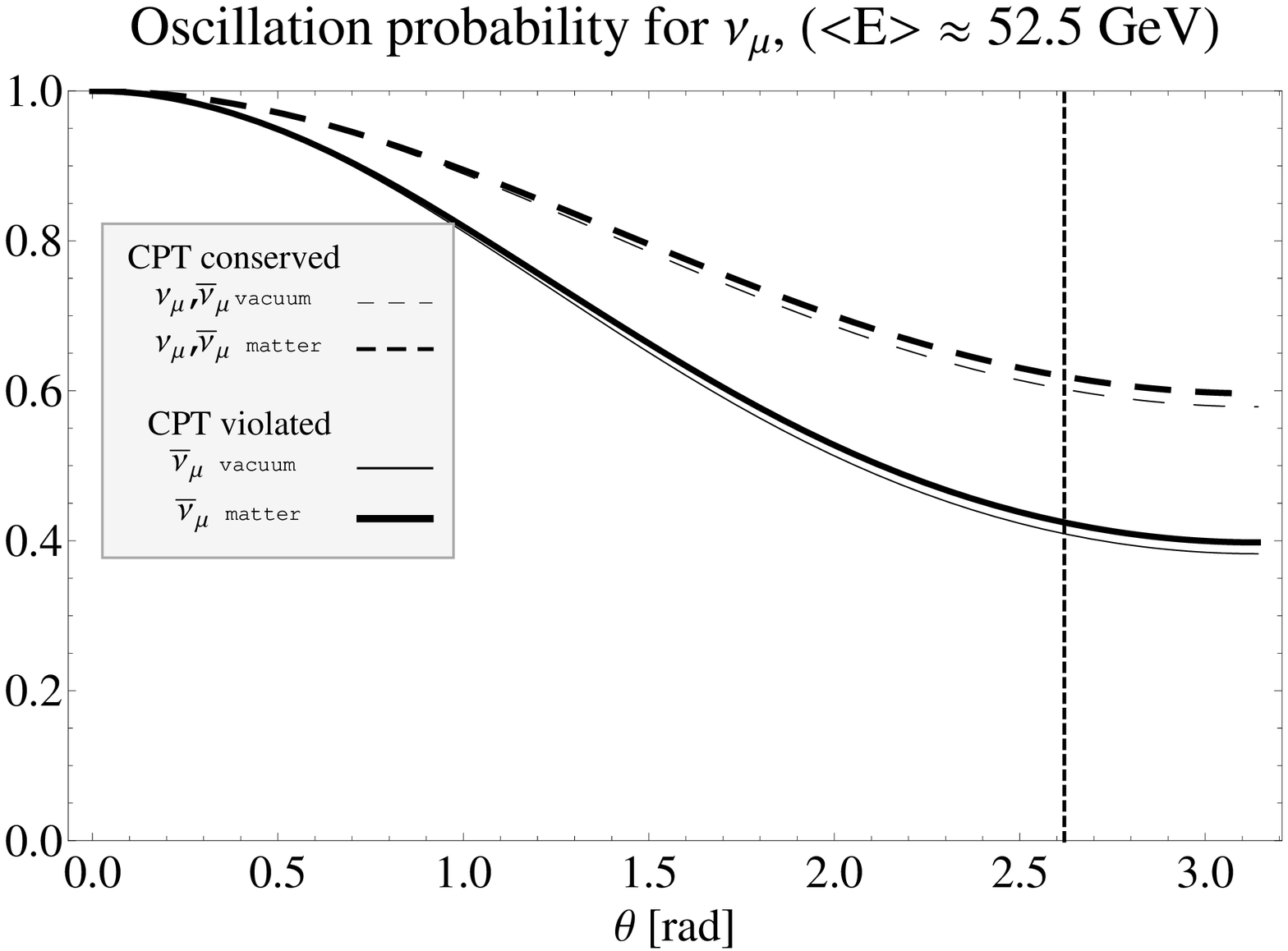,scale=0.18}
\epsfig{file=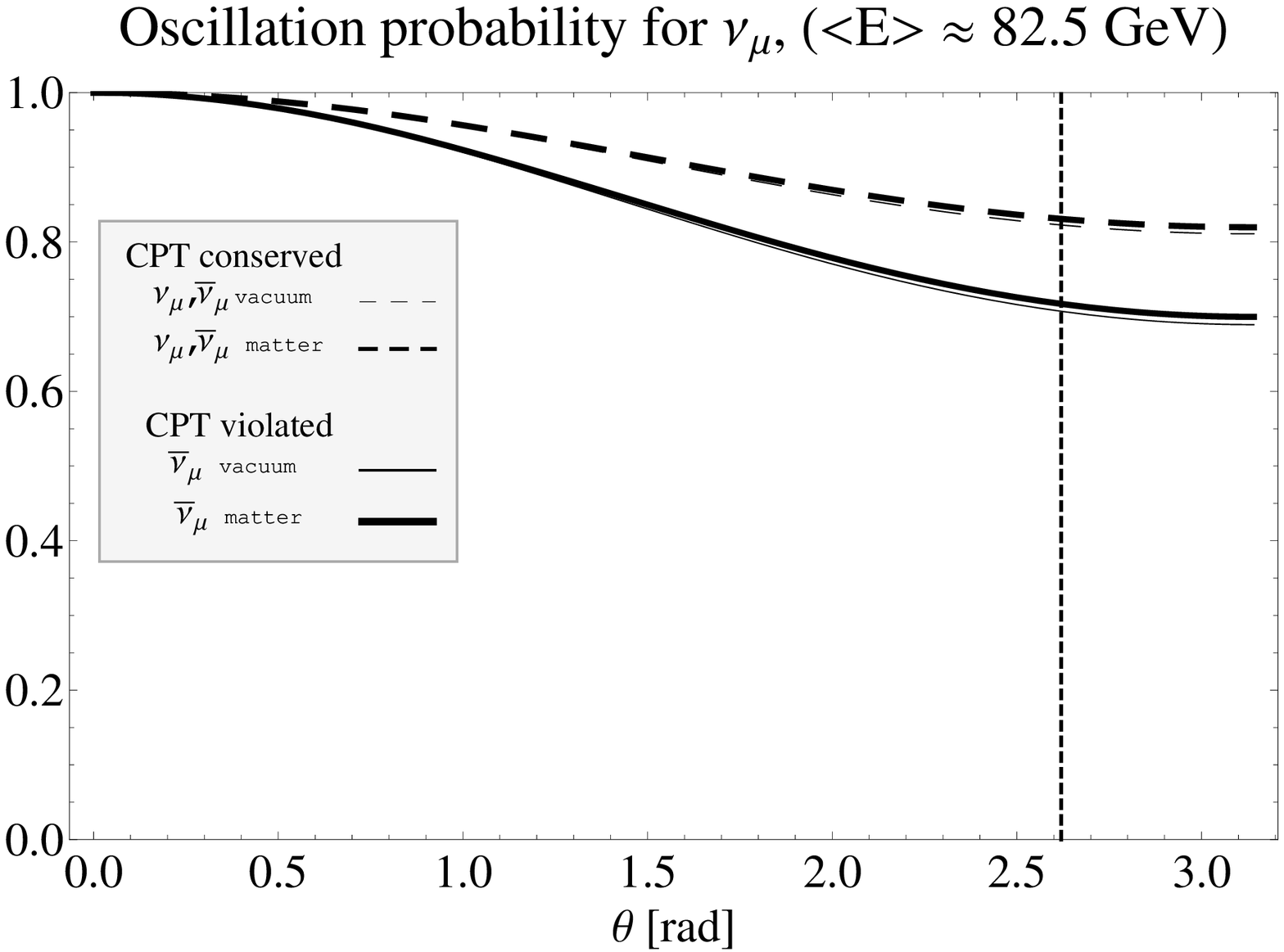,scale=0.18}
\epsfig{file=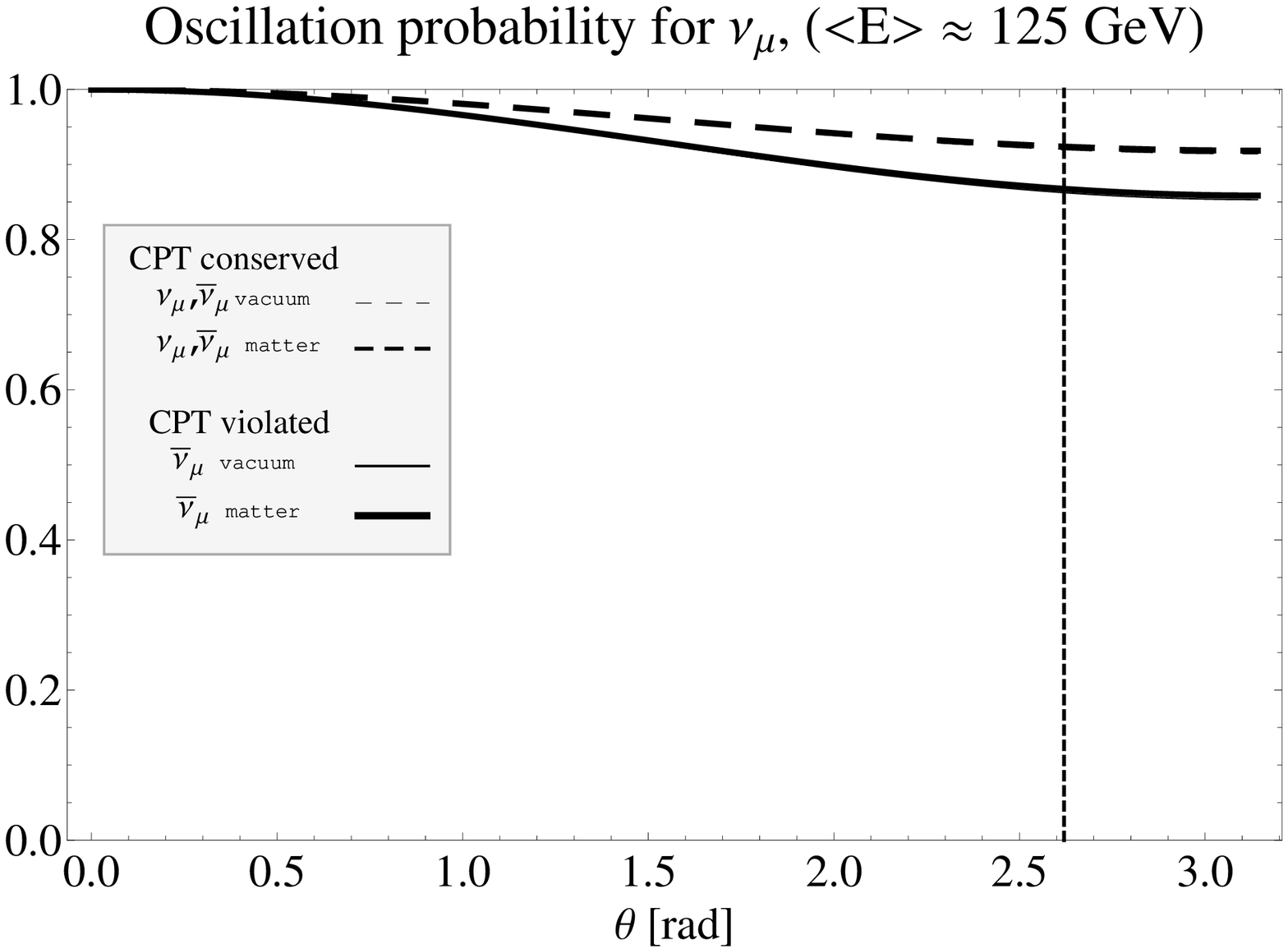,scale=0.18}
\epsfig{file=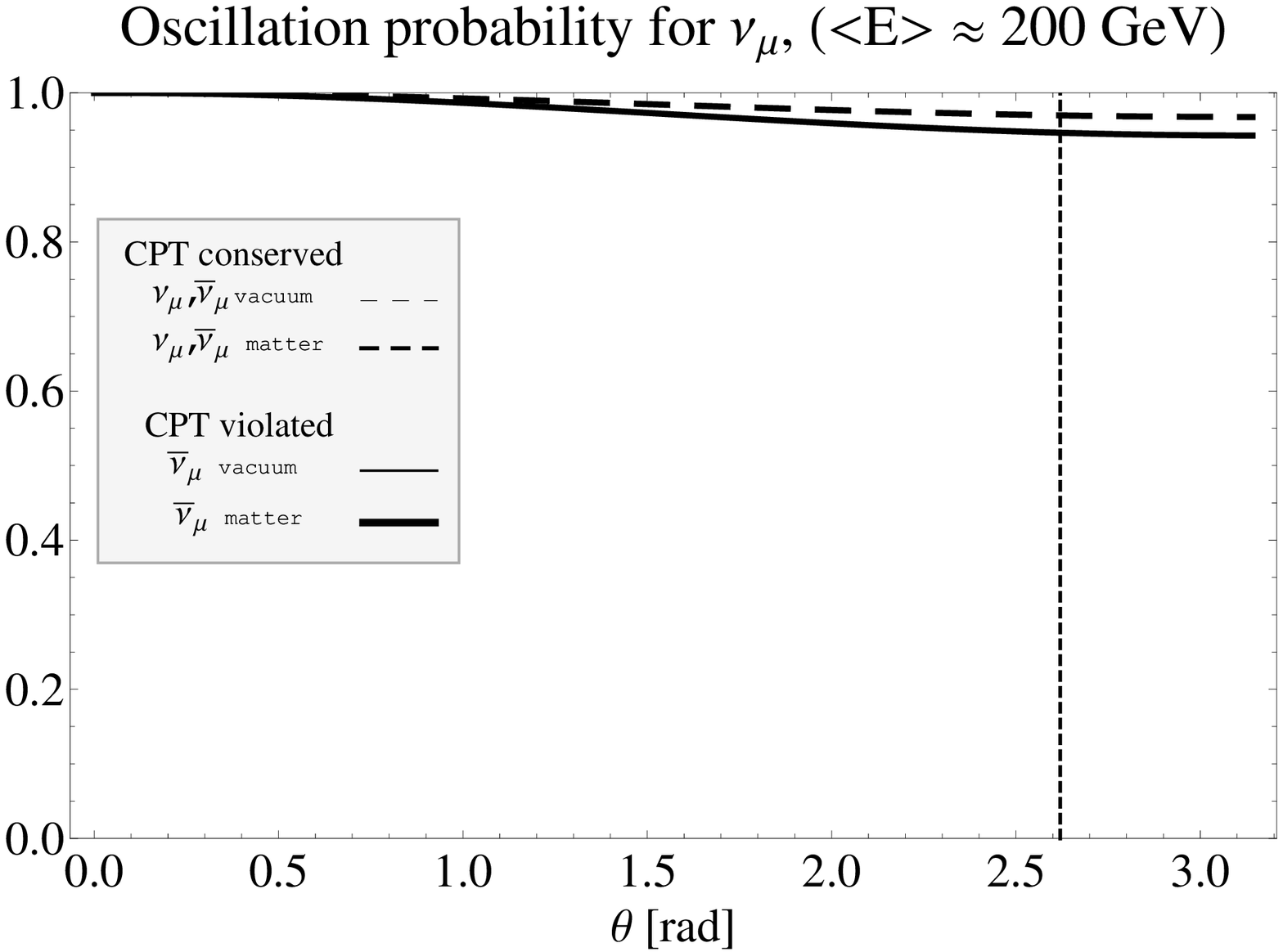,scale=0.18}
\caption{Eight different energy windows (corresponding to SK rate estimate above) for muon rates as a function of the arrival direction angle. Note the different suppression or enhancement of the oscillation in CPT violated (continuous) or conserved case (dashed) lines at vertical or quasi vertical arrival direction. These nearly vertical regions are above the vertical dashed line ($\pi - \theta \geq 2.6$ rad corresponding to $\theta \leq 33^{o}$) shown in each figure. The vertical downward North Pole points at $\theta \simeq \pi$ rad. The averaged energy label on the top of each figure corresponds to the channel region in figure above Fig. \ref{05}  }\label{613}
\end{center}
\end{figure}

 The calibrated muon rate figure \ref{05} shows in grouped channel graph number, the rate that we foresee following SK within a narrow vertical axis along each string. These prediction do not overlap  with previous one. In particular as we did mention we foresee a huge  rate of inclined events whose NC produce shower observable  by $3-4-5$ channel: this very rough estimate is based on the NC, and electron CC,NC shower : they  are well above $20.000$ NC  (for $\nu_{\mu}$ $\nu_{\tau}$) with additional thousands of shower by CC and NC  for $\nu_{e}$  and their antiparticles. The inclined tracks by nearly horizontal muons will excite the vertical string with a characteristic arrival time much shorter than any vertical shower event. Indeed the time difference in arrival for spherical shower along a string (each  DOM at $7$ m separation) is nearly $\Delta t_{0}\simeq t_{0}= h/c = 23 ns$; by triangulation any horizontal muon tracks and its Cherenkov  cone will record a quite shorter delay $\Delta t_{0}\simeq t_{0}\cdot(1-\frac{1}{\cos(\theta_{C}) \cdot n_{ice}}) \simeq 0.08 t_{0} = 1.84 ns$. Therefore this cluster of  event,  nearly coincident in time , might be a key test to calibrate the muon event rate in wide solid angle and possibly to meter the event rate at each channel.

\begin{figure}[htb]
\begin{center}
\epsfig{file=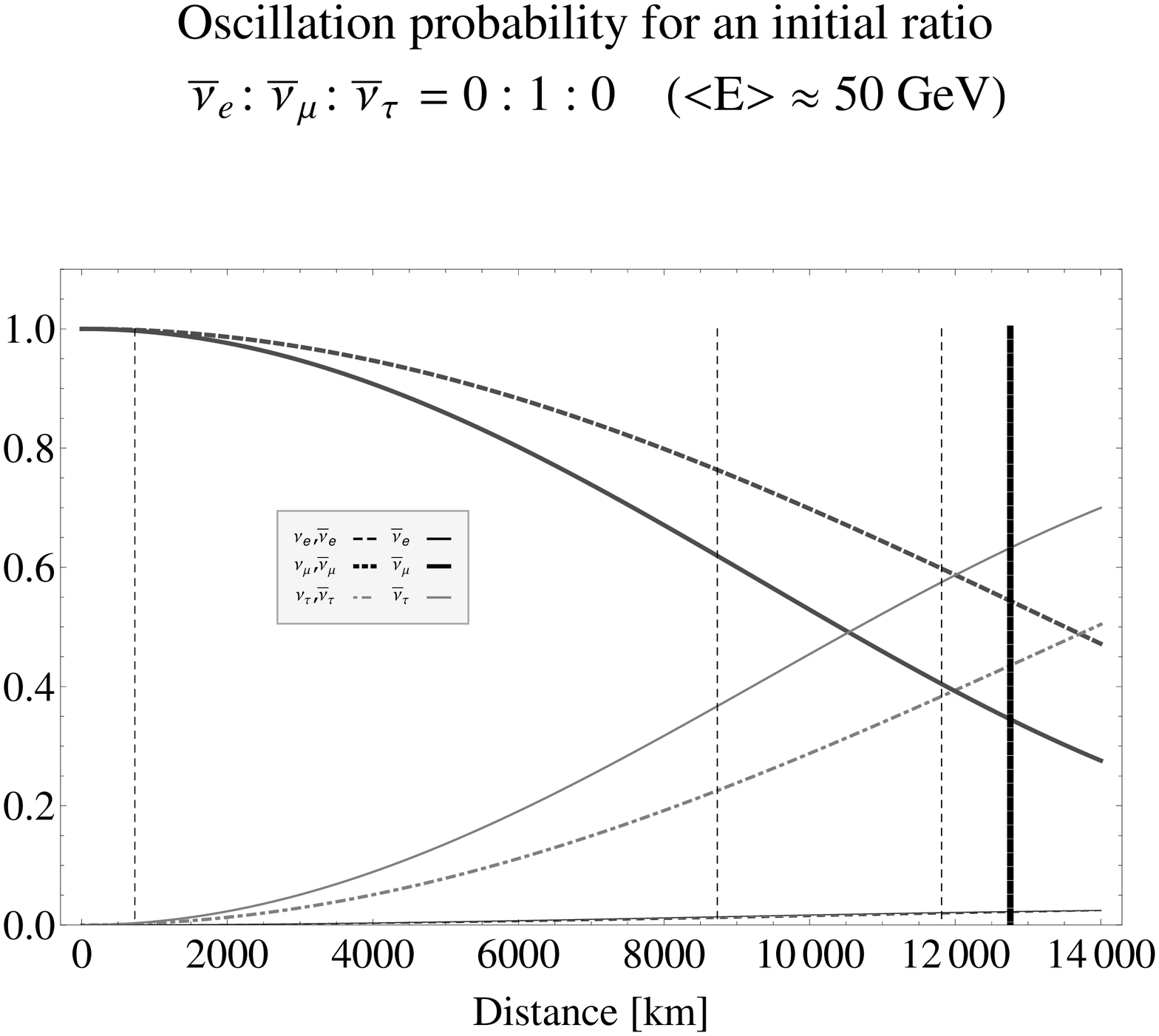,scale=0.25}
\epsfig{file=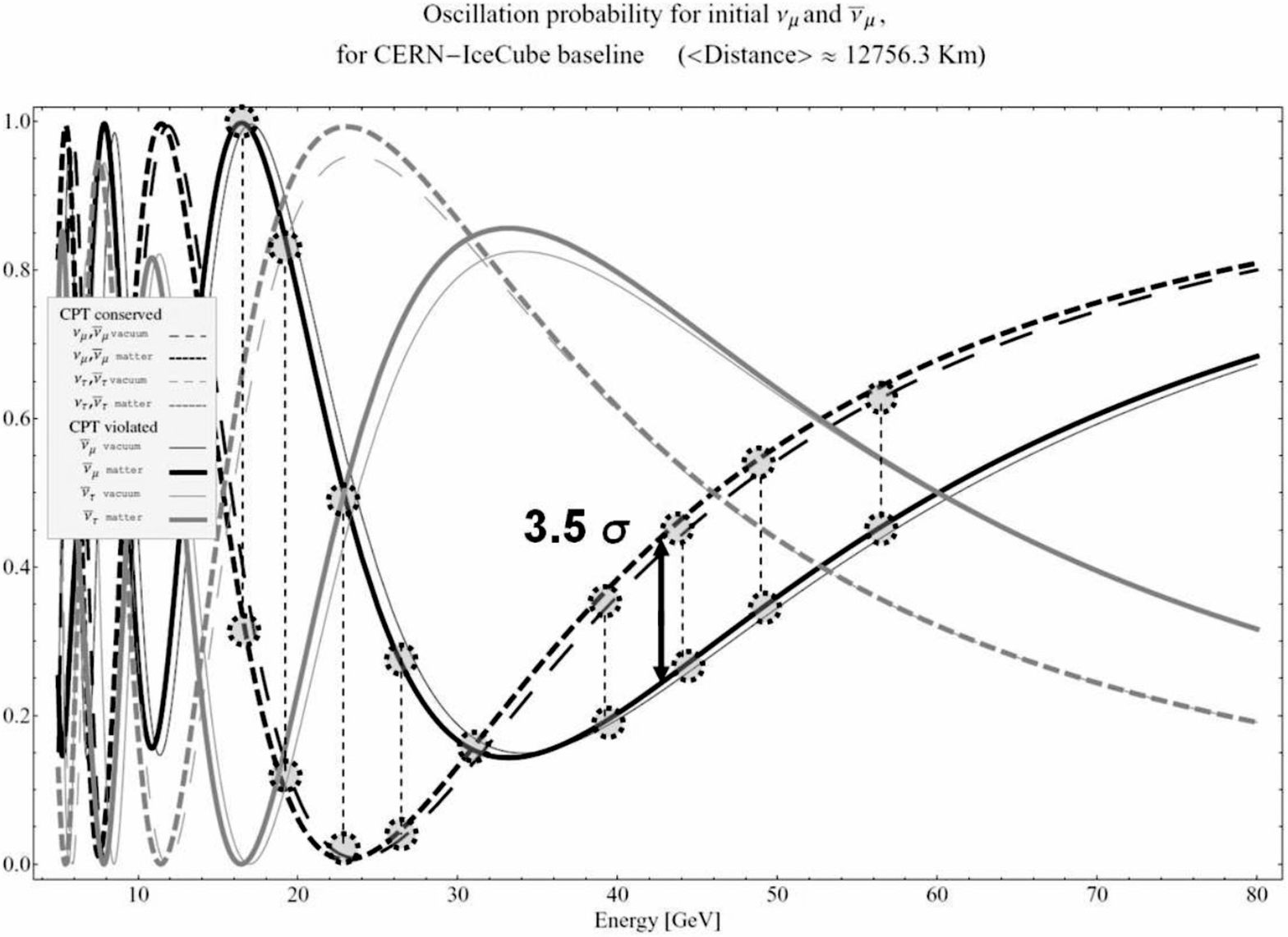,scale=0.13}
\caption{Left: In a CPT conserved (dashed lines) or in CPT violated case (continuous curve)  scenario the neutrino conversion($\nu_{\mu}\rightarrow\nu_{\tau} $) \emph{and the anti-neutrino} muon ($\overline{\nu}_{\mu}\rightarrow\overline{\nu}_{\tau} $), as well as survival
probability for given  $50$ GeV energy at different distances \cite{00}. The thick vertical line shows the Earth diameter, the dashed ones the Cern-Opera,SK,icecube distances. The same oscillation  in energy function and at earth diameter distance, in CPT conserved (dashed  line) and CPT violation (continuous line)  scenario reflects  additional suppression  of muon survival ($\overline{\nu}_{\mu}\rightarrow\overline{\nu}_{\mu} $)  at energies $ E_{\nu} \geq 44$ GeV. }\label{34}
\end{center}
\end{figure}

\clearpage

  \section{Conclusions}

  We did show the rate of atmospheric muon neutrinos along Deep Core string channels with some comments on the expected anisotropy. The main results are: a) a very sharp  peaked morphology in muon rate, namely a huge noise rate in low  range $3-6$ channel (ten thousand or above event a year) followed by b), a deep  minimum along channel  $7-9$ due to muon disappearance (contrasted partially by eventual CPT violation), whose rate maybe below $2000$ event a year. c) A global decay and  suppression of described events Fig (\ref{05}) in range above $10-50$, all along each channel, because of muon disappearance and additional anti-muon CPT violated suppression, by the vertical muon tracks that are more wide and they suffer  larger disappearance leading to  more  anisotropic behavior respect  to the averaged  SK rates. The present rate estimate differ from the most known ones \cite{05},{10},{20}; however the CPT violation influence foreseen in our previous paper \cite{00} plays the same role: to reduce the common muon survival, making more anti-tau appearance, from  channel above $\simeq 13$. The strong modulation by CPT violation at low channel ($3-6$) number,is quite remarkable, but it is nevertheless useless because a huge noise pollution by NC, electron CC,NC and nearly horizontal muon traces.

\end{document}